\documentclass[prb,11pt,showpacs,showkeys]{revtex4-1}

\usepackage{amsmath}    
\usepackage{txfonts}
\usepackage{graphicx}   
\usepackage{verbatim}   
\usepackage{color}      
\usepackage{subfigure}  
\usepackage{algcompatible}
\usepackage{newfloat}

\DeclareFloatingEnvironment[
    fileext=loa,
    listname=List of Algorithms,
    name=ALGORITHM,
    placement=tbhp,
]{algorithm}

\begin{document}

\title{Estimating the Number of Stable Configurations for the Generalized Thomson Problem}
\author{Matthew Calef}
\affiliation{Computational Physics and Methods, Los Alamos National Laboratory, Los Alamos, NM 87545-0001}

\author{Whitney Griffiths}
\affiliation{Financial Institutions Group, Bank of America, New York, NY 10036}

\author{Alexia Schulz}
\affiliation{Cyber Security and Information Sciences, MIT Lincoln Laboratory, Lexington, MA 02420-9108}

\date{23 March 2015}


\begin{abstract}
Given a natural number $N$, one may ask what configuration of $N$
points on the two-sphere minimizes the discrete generalized Coulomb
energy.  If one applies a gradient-based numerical optimization to
this problem, one encounters many configurations that are stable but
not globally minimal.  This led the authors of this manuscript to the
question, how many stable configurations are there?  In this
manuscript we report methods for identifying and counting observed
stable configurations, and estimating the actual number of stable
configurations.  These estimates indicate that for $N$ approaching two
hundred, there are at least tens of thousands of stable
configurations.
\end{abstract}


\maketitle

\section{Introduction}

``\emph{Computer trials indicate that in the range $70 \le N \le 112$,
the number of distinct configurations associated with each value of N
grows exponentially, i.e., $M(N) = 0.382 \, \exp(0.0497N)$. If
this trend is sustained for larger values of $N$, identifying global
minima among a large set of nearly degenerate states for complex
systems of this type will pose formidable technical challenges.}''
T. Erber and G. Hockney~\cite{ErbHock_Comment1}

\vspace{5mm}

For a natural number $N$, we denote by $\omega_N = \{{\bf
  r}_1,\ldots,{\bf r}_N\}$ any configuration of $N$ distinct points on
$\mathbb{S}^2$.  For a non-negative number $s$, one can ask what
configuration minimizes the energy
$$
E_s(\omega_N) := \sum_{i=1}^{N-1} \sum_{j=i+1}^N k_s(|{\bf r}_i - {\bf r}_j|),
\qquad\text{where}\qquad
k_s(r) = \left\{
\begin{array}{cc}
r^{-s} & \text{when $s>0$}\\
-\log r & \text{when $s=0$}.
\end{array}
\right.
$$ For $s=1$ this is known as the Thomson Problem~\cite{Thomson1}.  At
first glance this problem seems remarkably simple, yet there is not a
simple solution. In fact, Smale has identified a variant of this
problem as worthy of focus for this century~\cite{Smale1}.

The current theoretical progress is limited.  It is known that for any
$N$ and $s$ a globally minimal configuration exists. For a few special
cases of $N$ and $s$ there are rigorous proofs that certain
configurations are globally minimal. Finally, there are some
asymptotic estimates for the minimal energy as a function of $N$.  In
this last category P\'olya and Szeg\"o~\cite{PolyaSzego:1931}, using
measure-theoretic arguments, established some elegant estimates when
$s$ is less than the dimension of the set on which the problem is
posed, e.g., $2$ in the case of $\mathbb{S}^2$ (also
cf. Landkof~\cite[pp. 160-162]{Landkof1}).  Hardin and
Saff~\cite{HardinSaff1}, and Borodachov, Hardin and Saff~\cite{BHS1}
established similar results when $s$ is greater than or equal to the
dimension of the set in question.

While theoretical progress is difficult, the analyticity of $k_s$
makes this an inviting problem for numerical optimization,
particularly gradient-based optimizations.  Work in this area goes
back to at least 1977~\cite{MelnykKnopSmith:1977}, and there are two
efforts that particularly motivated this current effort. The first is
Erber's and Hockney's reports and
commentary~\cite{ErbHock2,ErbHock_Comment1,ErbHock1} on computational
experiments for the Thomson problem for $N$ up to 65 and $N$ up to
$112$, where they provide some initial estimate for the growth in the
number of stable configurations as a function of $N$.  This work also
includes estimates for the first two terms in the asymptotic expansion
for the minimal energy.  The first term is in agreement with the
earlier work of P\'olya and Szeg\"o, and the second term was later
identified in a formal conjecture by Kuijlaars and
Saff~\cite{KuijlaarsSaff1} and later generalized to a large class of
two-manifolds~\cite{Calef:2013}.  The second effort is the work by
Wales and Ulker~\cite{WalesUlker:2006}, and the work by Wales, McKay
and Altschuler~\cite{WalesMcKayAltschuler:2009} that led to the
Cambridge Cluster Database, which reports, for many $N$ and $s=1$, the
lowest known energy for the Thomson problem.

A significant challenge for numerical optimization is that many
configurations are locally minimal, i.e. stable, with respect to
$E_s$, but not globally minimal.  This motivated us to attempt to
answer the question: how many stable configurations for a given $N$
and $s$ are there?  An earlier work that answers a similar question is
that of Hoare and McInnis, who identify the distinct stable clusters
of modest numbers of point-particles interacting through Lennard-Jones
and Morse potentials~\cite{HoareMcInnis:1976}.  For the Lennard-Jones
potential the number of stable clusters grew rapidly.  The present
work estimates this growth for the generalized Thomson problem for $N$
up to $180$ and for $s=0$, $1$, $2$, and $3$, and reports some methods
we found useful.

A central question is whether the number of distinct local minima within the
energy landscape grows exponentially with the number of points.
Stillinger and Weber present the following informal argument
suggesting an affirmative answer~\cite[p. 980]{StillingerWeber:1982}.
If one can convert one stable configuration into another with changes
that are localized in space to within a fixed number nearest neighbor
lengths, then, as the number of points grows, so should the number of
available independent changes, and the number of stable configurations
grows will grow exponentially. If the number of stable configurations
does not increase exponentially with $N$, then this would suggest that
changes from one stable configuration to another cannot be
accomplished with only localized changes.

Our work began by generating a large library of stable configurations,
and we describe our methods in Section~\ref{sec:generate_stable}.  In
doing this, we found that our optimization program would find
rotations and reflections of the same stable configuration.  In
response, we used graph-isomorphisms of the Delaunay Triangulation as
a means to recognize quickly a particular stable configuration.  This
method accelerated our work considerably, but there are some subtle
ways it can fail.  In particular we found two distinct stable
configurations whose Delaunay Triangulations share the same graph
structure.  This is described in Section~\ref{sec:graph_iso}.  Even
after months of numerical experiments running on many compute cores,
the fraction of the most recent experiments that generated new
configurations never dropped to zero, making it clear that there are
many stable configurations we did not see.  The problem of estimating
the number of configurations we didn't see is an example of the
broader ``unseen species'' problem, which arises in many settings such
as linguistics and ecology.  We apply a method developed partly by
linguists to provide estimates in Section~\ref{sec:unseen} for the
actual number of stable configurations.

\section{Stable Configurations} \label{sec:generate_stable}

\subsection{Optimization}

We used an iterative unconstrained optimization strategy described in
Section III.A of our prior work\cite{Calef:2013} to generate candidate
stable configurations.  The method consists of non-linear conjugate
gradient (NLCG) with line minimization and, when that method no longer
made progress, Newton's Method.  Our experience is that NLCG with line
minimization was most effective up until near the end of the
optimization.  Near the end of the optimization calculation,
presumably when the configuration is at a point where the objective
function $E_s$ is locally quadratic, Newton's Method would often make
progress when NLCG could not.

When computing the energy, $E_s(\omega_N)$, one has roughly $N^2/2$
summands that vary widely in range, and a direct summation can lead to
roundoff errors.  We controlled for this error by logarithmically
binning our summands and only adding the content from the same bin.
This allowed us to ensure that we never added two numbers whose ratio
was more than two or less than one half, until the end when we summed
the contents of the bins from lowest to highest.  Because we could
bound the error for summation in a given bin, and because we could
count the number of summations in the bin, we were able to estimate
the error in our sums.  This approach follows the work of
Higham~\cite{Higham:1993} and Demel and Hida~\cite{DemmelHida:2003}.

\subsection{Testing for Stability}

Given a candidate stable configuration, we use the criteria described
in Section III.B of a previous publication~\cite{Calef:2013} to test
for stability.  The central assumption in this criteria is that our
iterative optimization strategy will produce a candidate configuration
$\omega^c_N$ that is close enough to an actual stable configuration
$\bar \omega_N$, so that the gradient at $\bar \omega_N$, which is
zero, may be expressed as a linear expansion of the gradient about
$\omega^c_N$.  That is
\begin{equation}\label{eq:linear}
0 = \nabla E_s (\bar \omega_N) \approx \nabla E_s(\omega^c_N) + 
\nabla^2E_s(\omega^c_N)(\bar \omega_N - \omega^c_N),
\end{equation} where $\nabla E_s$ and $\nabla^2 E_s$ are the gradient and the
Hessian respectively of the objective function with respect to the
$2N$ angular free parameters.  If this approximation were exact, it
would allow us to bound the term $\bar \omega_N - \omega^c_N$, where
subtraction is applied to the $2N$-dimensional space of
configurations.  Conceptually the calculation is
$$
-\nabla^2 E_s(\omega^c_N)^{-1}\nabla E_s(\omega^c_N) = (\bar \omega_N - \omega^c_N),
$$
$$
\left\|\nabla^2 E_s(\omega^c_N)^{-1}\right\|_2
\left\|\nabla E_s(\omega^c_N)\right\|_2 
\geq
\left\|\bar \omega_N - \omega^c_N\right\|_2
\geq
\left\|\bar \omega_N - \omega^c_N\right\|_\infty.
$$ Here $\|\cdot\|_2$ is the unnormalized two-norm allowing the
bound of the infinity-norm.

The Hessian is not invertible, however.  For our choice of coordinates
there are three rotations of the sphere that do not change the
relative distance between the points and hence don't change the
energy.  While there are choices of coordinates free of such rigid
rotations, those coordinate systems degraded the performance of NLCG.
Consequently the three lowest eigenvalues of the Hessian are zero.
The gradient has no projections along the corresponding eigenvectors,
and we may choose a rotation of $\bar \omega_N$ so that the difference
$\bar \omega_N - \omega^c_N$ similarly does not project along these
eigenvectors.  We let $\lambda^*_\text{min}$ denote the fourth lowest
eigenvector of the Hessian and then we have
$$
\frac{\|\nabla E_s(\omega^c_N) \|_2}{\lambda^*_\text{min}} 
\geq
\left\|\bar \omega_N - \omega^c_N\right\|_2
\geq
\left\|\bar \omega_N - \omega^c_N\right\|_\infty.
$$ Change in angle on the sphere bounds from above change in position,
and so $\left\|\bar \omega_N - \omega^c_N\right\|_\infty$ provides a
bound on the distance between corresponding points in the
configurations $\bar \omega_N$ and $\omega^c_N$.

Our criteria for stability is that 
\begin{equation}\label{eq:stab_crit}
\frac{\|\nabla E_s(\omega^c_N) \|_2}{\lambda^*_\text{min}} 
\le
\frac{\min_{{\bf r}_i\ne{\bf r}_j\in \omega^c_N}|{\bf
    r}_i - {\bf r}_j|}{10,000},
\end{equation} which, in conjunction with the assumption that error in the
approximation in Eq.~\eqref{eq:linear} is negligible, leads to the
conclusion that no point in $\omega^c_N$ is further from the
corresponding point in $\bar \omega_N$ by more than one ten-thousandth
the minimum pairwise separation of the points in $\omega^c_N$.  An
important consequence of this is that if two configurations satisfy
Eq.~\eqref{eq:stab_crit}, and if there is a rotation and reflection
that aligns them to within one five-thousandth of both of their
minimum pairwise distances, then, we say that they are instances of
the same stable configuration.  As previously noted~\cite{Calef:2013}
this criteria relies on bounding the infinity-norm with the
unnormalized $2$-norm.  Such a bound is tight only when all the
components except one are zero.  The implication is that the maximum
difference between a point in our candidate configuration and a true
stable configuration is likely considerably less than one
ten-thousandth of the minimum pairwise separation of the points within
the configuration in question.  For candidate configurations that we
believed were instances of the same stable configuration, we could
often align them to greater accuracy.

The condition in Eq~\eqref{eq:stab_crit} is a useful, reasonably
motivated, heuristic for marking a configuration as stable.  More
rigorous bounds would require estimating the error in
Eq.~\eqref{eq:linear}.

\section{Delaunay Triangulations and Graph Isomorphisms} \label{sec:graph_iso}

Since the energy $E_s$ depends only on the distances between points,
it is invariant under isometry.  However, that two configurations of
points have the same energy does not ensure that there is an isometry
between the two configurations.  With this in mind we only called two
configurations the same if we could find an isometry that mapped one
configuration onto the other to within the tolerances described in the
previous section.

This leaves the question of how to search for an isometry between two
configurations of similar energy, which we'll denote here as
$\omega_N^1 = \{{\bf s}_1,\ldots,{\bf s}_N\}$ and $\omega_N^2 = \{{\bf
  r}_1,\ldots,{\bf r}_2\}$.  A simple approach is to apply
Algorithm~\ref{alg:simple} described in this manuscript. While there
are some optimizations such as, at line 5, first testing that $|{\bf
  r}_i - {\bf r}_j| = |{\bf s}_1 - {\bf s}_2|$, this algorithm is
expensive and must be applied to every pair of configurations with
similar energy.

\begin{algorithm}
\caption{\label{alg:simple}A simple way to search for isometries
  between $\omega_N^1$ and $\omega_N^2$.}
\begin{algorithmic}[1]
\STATE {Isometry Found $\gets$ False.}
\STATE {$\varepsilon \gets 2\min \left\{ 
\frac{\min_{{\bf s}_i\ne{\bf s}_j\in \omega^1_N}|{\bf s}_i - {\bf s}_j|}{10,000}, 
\frac{\min_{{\bf r}_i\ne{\bf r}_j\in \omega^2_N}|{\bf r}_i - {\bf r}_j|}{10,000}
\right\}$}
\FOR{ ${\bf r}_i \in \omega_N^2$ }
\FOR{ ${\bf r}_j \in \omega_N^2 \backslash \{{\bf r}_i \}$  }
\IF {there is a rotation of $\omega_N^2$ so that ${\bf r}_i = {\bf s}_1
\in \omega_N^1$ and so that ${\bf r}_j = {\bf s}_2 \in \omega_N^1$ to within $\varepsilon$} 
\IF { this rotation is such that $\left\|\omega_N^1 - \omega_N^2\right\|_\infty < \varepsilon$ } 
\STATE { Isometry Found $\gets$ True. } 
\ELSE 
\STATE {$\tilde \omega_N^2 \gets $ the reflection of the rotation
  of $\omega_N^2$ about
  the plane defined by ${\bf s}_1$, ${\bf s}_2$ and ${\bf 0}$.}
\IF { $\left\|\omega_N^1 - \tilde \omega_N^2\right\|_\infty < \varepsilon$ }
\STATE { Isometry Found $\gets$ True. } 
\ENDIF
\ENDIF
\ENDIF
\ENDFOR
\ENDFOR
\end{algorithmic}
\end{algorithm}

\subsection{Delaunay Triangulations}

We found a more effective algorithm was to look for isomorphisms
between the graphs formed from the extremal edges in the Delaunay
Triangulations of the configurations in question.  For brevity we
shall refer to this as an extremal triangulation.  Essentially we are
looking at the edges in the set of triangles that make up the surface
of the smallest polyhedron containing a configuration, $\omega_N$.

\begin{figure}
\begin{center}
\includegraphics[height=7cm]{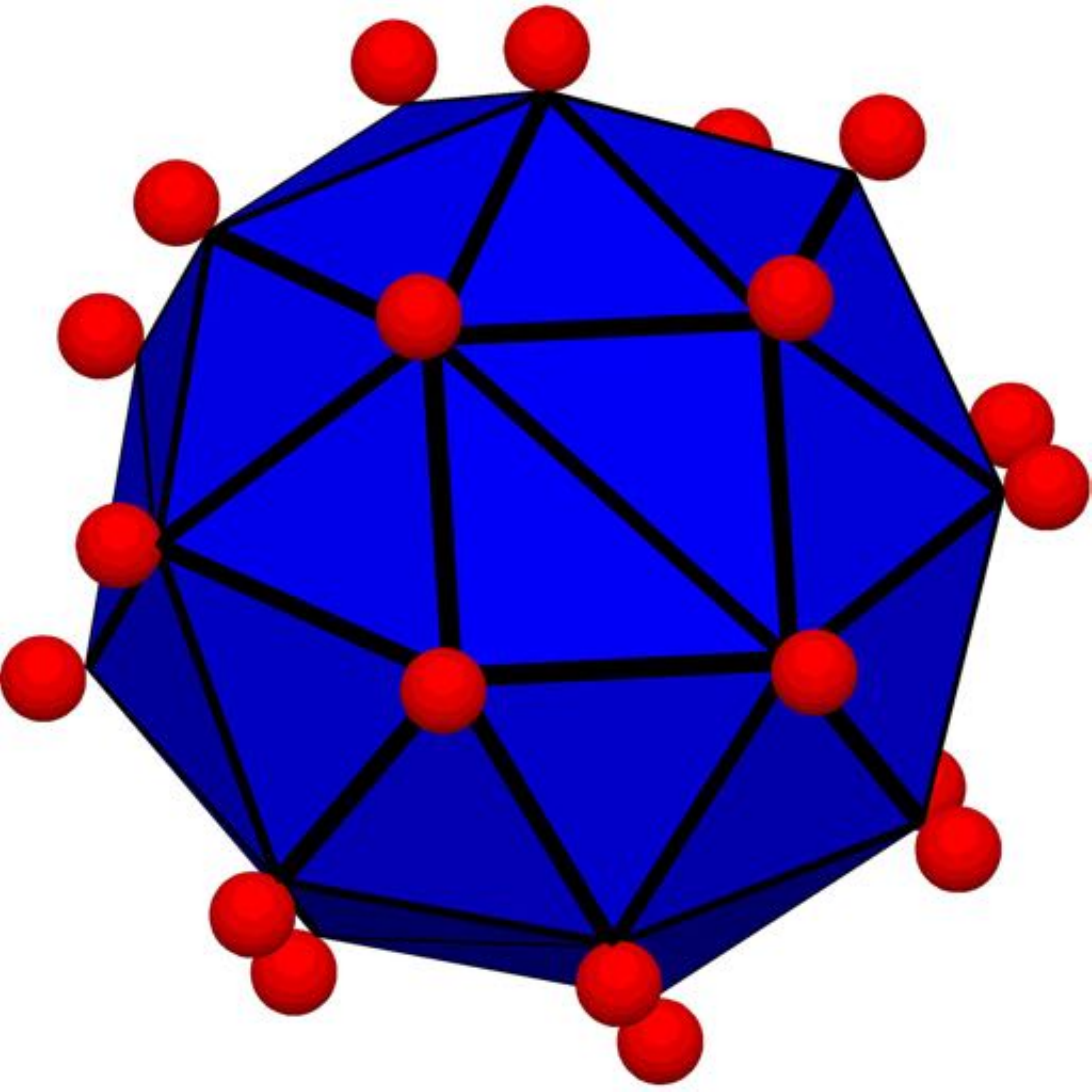}
\end{center}
\caption{\label{fig:sph24} This is one of many possible extremal
  triangulations for this configuration of $24$ points.}
\end{figure}

To compute the extremal triangulation we used the QHULL software
package~\cite{QHull:1996}.  One immediate observation was that certain
configurations did not have unique extremal triangulations, for
example, the configuration with the lowest observed energy for $N=24$
and $s=1$, shown in Figure~\ref{fig:sph24}.  The four points displayed
toward the middle of the image are the vertices of a square, and
either diagonal can be part of a valid extremal triangulation.
Because there can be degenerate extremal triangulations, the
assumption that distinct extremal triangulations indicate
non-isometric configurations is not, in general, correct.  A simple
test for non-degeneracy is to compute the set of unit normal vectors
for the extremal faces, and to make sure that the dot-product of any
two is bounded away from one.

In the case that a configuration has a non-degenerate extremal
triangulation, the edges of the triangulation and the points in the
configuration form a graph, and this graph is invariant under rotation
and reflection of the underlying configuration.  In the degenerate
case, a rotation or reflection may lead QHULL, due to round-off
errors, to find a different, but equally valid, extremal
triangulation.

\subsection{Graph Isomorphisms}

A graph on a sphere is a planar graph in that, by choosing one face to
be mapped to the unbounded component of the plane, the graph can be
mapped onto the plane.  In doing this the edges that bound this face
are retained, and the graph structure is preserved whether the graph
is embedded on $\mathbb{S}^2$ or $\mathbb{R}^2$.  There are efficient
algorithms to determine isomorphisms of planar graphs and we use one
following the work of Lins~\cite{Lins:1980}.  The approach is to
generate a tag for each graph with the property that two graphs are
isomorphic if, and only if, the two tags are the same.  The cost for
finding isomorphisms between $M$ instances of graphs, or for finding
isometries between $M$ configurations, can be written as
$$
C_1 M + C_2 M^2.
$$ In our approach $C_2$ is the cost of a searching for matching tags,
i.e. string comparisons, while $C_1$ is the cost of generating the tag.
For large $M$, this has substantial benefits, over the case that $C_1$
is zero, but $C_2$ is the cost associated with
Algorithm~\ref{alg:simple}.

The specific method we use for generating a tag is given in
Algorithm~\ref{alg:tag}.  We denote our graph as a set of vertices $V$
and a set of edges $E$, and for any $v\in V$ we denote by $E(v)$ the
set of edges that have $v$ as an endpoint.  To each vertex $v$ we
assign a natural number $i_v$.  The central idea in
Algorithm~\ref{alg:tag} is to search for the lexically lowest encoding
of a representation of the connectivity matrix, where we are searching
over a set of possible orderings of vertices.  We used an MD5 hash of
the connectivity matrix simply to use less memory.  The requirement
that the graph be planar is what allows us to generate the unique
ordering of $W$ at line twelve. We stored the configuration with the
ordering of points that generated the lexically lowest encoding,
i.e. the tag.

\begin{algorithm}
\caption{\label{alg:tag} Generating a tag for a planar graph.}
\begin{algorithmic}[1]
\STATE { Tag $\gets$ None. }
\FOR { $v\in V$ } 
\FOR { $e\in E(v)$ }
\FOR { $r\in \{\text{Clockwise},\text{Counterclockwise}\}$ }
\STATE { Reset all indices $i_v$ for $v\in V$. }
\STATE { $i_v \gets 1$. }
\STATE { $i_w \gets 2$, where the edge $e$ joins the points $v$ and $w$. } 
\STATE { $n \gets 3$.}
\WHILE { there is a vertex that has not been indexed } 
\STATE { $x \gets $ the vertex with the lowest index that has an
  unindexed neighbor.}
\STATE { $y \gets x$'s neighbor with the lowest index. } 
\STATE { $W \gets $ set of neighbors of $x$
 ordered by $r$ and starting with $y$.}
\FOR { $w \in W$}
\IF {$w$ has not been indexed }
\STATE { $i_w \gets n$. } 
\STATE { $n \gets n+1$. }
\ENDIF
\ENDFOR
\ENDWHILE
\STATE { $P \gets $ the connectivity matrix of the graph where the
  vertices are ordered by their indexing.}
\STATE { $T \gets $ the MD5 cryptographic hash of $P$. }
\IF { Tag $=$ None or $T <$ Tag }
\STATE { Tag $\gets T$. }
\ENDIF
\ENDFOR
\ENDFOR
\ENDFOR
\end{algorithmic}
\end{algorithm}

When we generated a new stable configuration for a given $N$ and $s$
we would, when the new configuration had a non-degenerate extremal
triangulation, also generate the associated tag.  We would then
collect the already generated configurations for that $N$ and $s$
whose energies were close to the energy of the newly generated
configuration.  Within the subset of these with unique extremal
triangulations, we would search for the tag associated with the new
configuration.  If we found it, then because we stored the
configurations with the orderings of points that generated the tag, we
knew the rotation and reflection necessary that would be the isometry.
It was our experience that, when there was an isometry, this method
found it immediately.  Further, it was our experience that almost all
of the configurations had non-degenerate extremal triangulations.

If we did not find matching tags, then we used
Algorithm~\ref{alg:simple} to search for an isometry between the newly
generated configuration and the existing configurations with similar
energies.  We only characterized a configuration as new for a given $N$
and $s$ if every configuration with that $N$ and $s$ had an energy
that was sufficiently different to ensure that there was no isometry
or that the application of Algorithm~\ref{alg:simple} did not find an
isometry.  The graph-isomorphism technique sped the process of finding
isometries when they existed, but it was never used by itself to
determine if a configuration was new or isometric to an existing one. 

There are two reasons why one should not rely exclusively on
isomorphisms of non-degenerate extremal triangulations.  The first is
that it is possible, although we didn't see this case, that the graph
has a non-trivial automorphism, but that the associated mapping of
points is not a self-isometry.  Put another way, there may be two
orderings of the points that lead to the same lexically minimal tag.
An indication of this would be that, at line twenty-two of
Algorithm~\ref{alg:tag}, Tag was not None and $T = \text{Tag}$, but,
that there is no isometry between the configurations that preserves
the orderings of points.  An extremely simple example is a triangle
where no two sides have the same length. It has no self-isometries,
but six graph-automorphisms.  Also, there is the remote possibility
for collisions in the MD5 algorithm.

\begin{figure}
\begin{center}
\includegraphics[height=7cm]{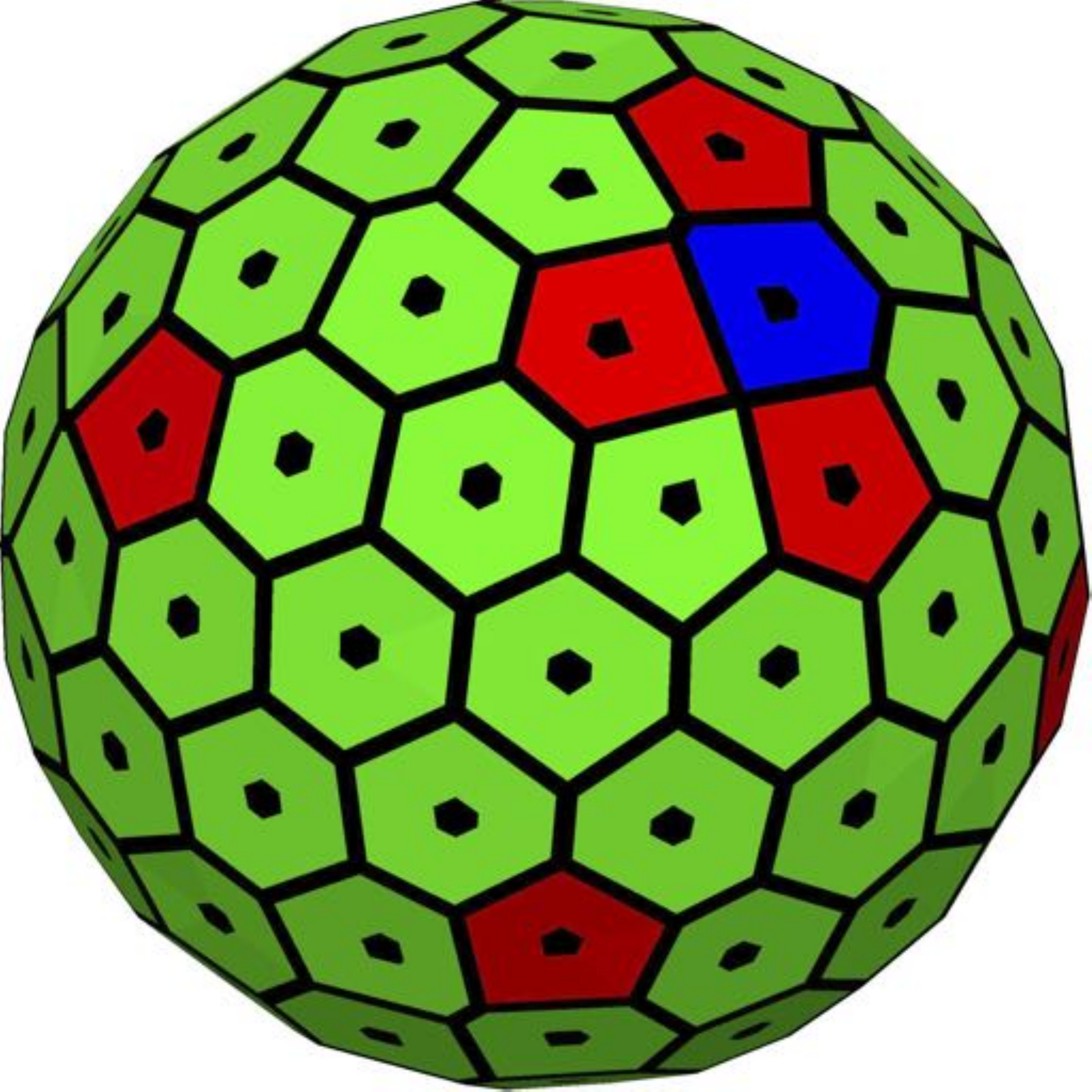}
\includegraphics[height=7cm]{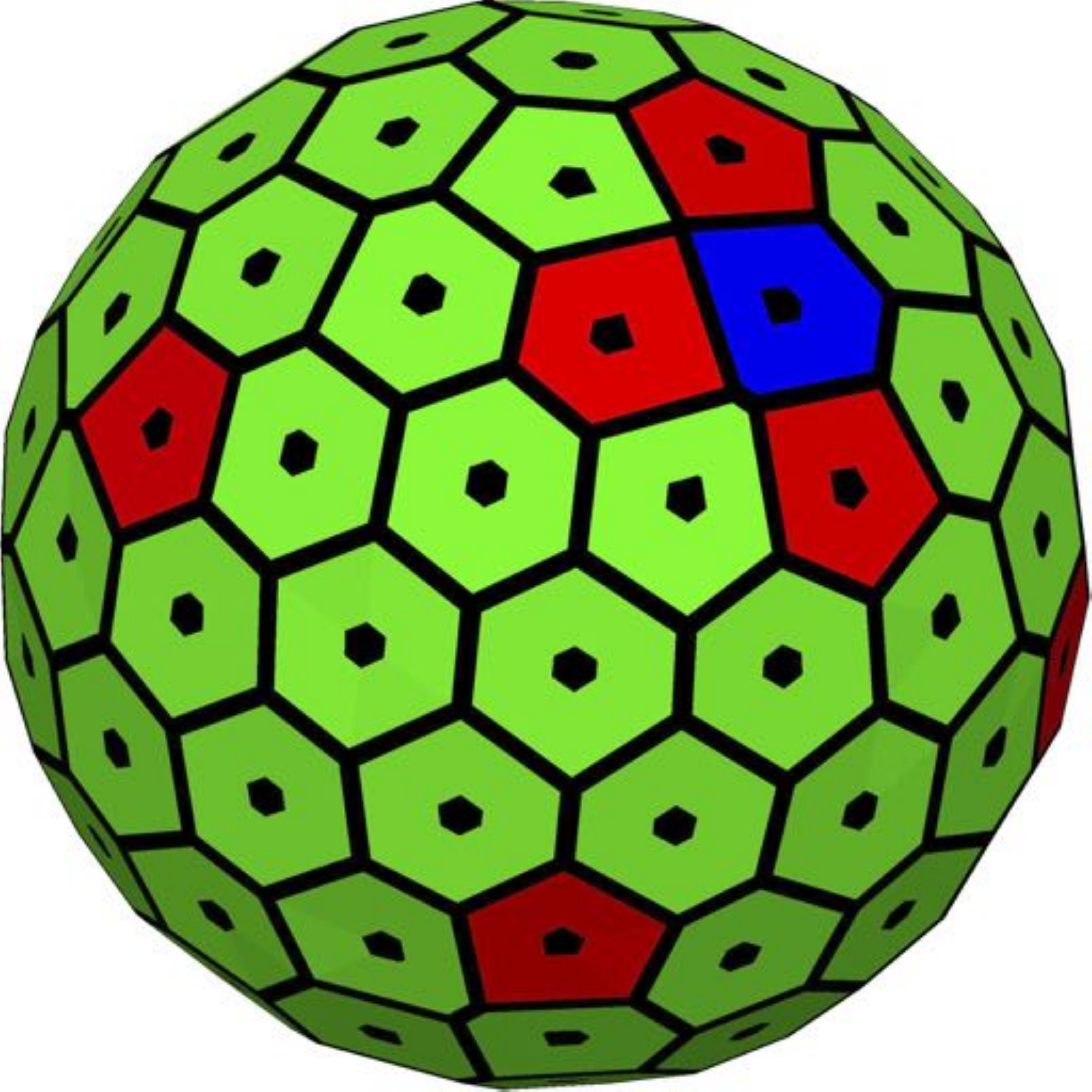}
\end{center}
\caption{\label{fig:n102_s2} On the left is the configuration with the
  fourth lowest energy for $N=102$ and $s=2$, on the right is the
  configuration with the fifth lowest energy.  They have the
  non-degenerate extremal triangulation with isomorphic graphs, but
  are distinct stable configurations.  The dark (blue in the online
  version) cell has seven edges, two of which in the upper left and
  lower left are extremely short.  The reader will notice that these
  two edges differ in the image on the left and the right.}
\end{figure}

The more significant reason that graph isomorphisms alone are not
sufficient to identify stable configurations is that we found two
distinct configurations that both have non-degenerate extremal
triangulations, but whose graphs were isomorphic.  For $N=102$ and
$s=2$, the configurations with the fourth and fifth lowest energy have
non-degenerate extremal triangulation with isomorphic graphs.  The
dual graphs, i.e. the Voronoi cells, are shown in
Figure~\ref{fig:n102_s2}.  The difference in energy is substantially
more than the estimated error in the energy sums.  The energies are
$5582.2331644897$ and $5582.2332117851$
respectively. Algorithm~\ref{alg:simple} did not identify an isometry.
Further, out of thousands of computer trials, we reproduced the fourth
lowest configuration $205$ times and the fifth lowest configuration
$100$ times -- these were not rare configurations.

\section{Unseen Species}\label{sec:unseen}

After months of running on many compute cores, we found that for most
$N$ between $120$ and $180$ the rate at which we discovered new stable
configurations was still far from zero.  We took this as an indication
that more trials would result in more distinct stable configurations,
and that we had not seen all of them.  In response, we aimed to
estimate the number of stable configurations that we didn't see in our
trials.  Such an estimate cannot be made without additional
assumptions, which we shall make clear as we proceed.  There is some
precedent for trying to estimate the number of unseen species. For
example, Efron and Thisted estimated the number of words Shakespeare
knew~\cite{EfronThisted:1976}, although their approach is more
sophisticated than ours.

In broad terms our approach is as follows: We first compute the
Good-Turing Frequency described below.  This is an estimate for the
combined probability of all the configurations we did not see.  In
addition this method produces estimated probabilities for the $S$
stable configurations we did see, $\{p_1,p_2,\ldots,p_S\}$.  We assume
that when these estimated probabilities are sorted in decreasing order
the tail has a certain analytic form, i.e. that there is a $p(n)$ so
that $p_n = p(n)$ for large $n$. We obtain $p(n)$ from the data, and
use it to compute how many more configurations we would need for the
sum of the probabilities of those unseen configurations to agree with
the Good-Turing Frequency.

The first assumption is that the number of stable configurations is
finite.  While this seems intuitively true, the function $f(x) = x
\sin(1/x)$ for $x\ne 0$ and 0 for $ x = 0$ has infinitely many local
minima on the closed unit interval, indicating that a proof that there
are finitely many stable configurations will depend on domain specific
information.

A second assumption is that, were we to use a different gradient based
optimization technique, the estimated probabilities for configurations
we observed wouldn't be so different as to change dramatically the
estimates for the unseen species. While we have no proof, our
instincts are that the basins of attraction for gradient descent
methods all are qualitatively the same, and that the initial random
configurations were sufficiently disordered so as not to be ``nearer''
a particular subset of stable configurations.  Indeed, efforts to
avoid the preponderance of stable configurations while searching for
the global minimum has lead researchers away from purely
gradient-based methods such as the work by Morris, Deavon and
Ho~\cite{MorDevHo1}, and the work by Lakhab and
Bernoussi~\cite{Lakhbab:2013}.

We now briefly summarize I. J. Good's description of the Good-Turing
estimate~\cite{Good:1953}.  Suppose we perform $T$ trials where we
observed some number of distinct species -- stable configurations in
our case.  We let $n_r$ denote the number of species that we observed
$r$ times, and so
\begin{equation}\label{eq:sp}
\sum_{r=1}^\infty n_r r = T.
\end{equation} We then ask, for a species that we saw $r$ times, what is a
reasonable estimate for the fraction of the population that consists
of that species?  The most straightforward estimate, $r/T$, has the
drawback that the sum of the fractions is one, i.e. this estimates
assumes that there were no unseen species.  This is almost certainly
wrong in our case.  If the likelihood of seeing each species is
described by a binomial distribution, which is reasonable in our case,
Good arrives at the following estimate~\cite[\S2 Eq. 15]{Good:1953}
for the probability of a species that was observed $r$ times
$$
\frac {r+1}{T+1} \frac {{\mathcal E}_{T+1}(n_{r+1}) } {{\mathcal E}_T(n_r) }.
$$ Here, in a slight abuse of notation, ${\mathcal E}_T(n_r)$
indicates the expectation value for the number of species that we
would expect to see $r$ times in $T$ trials.  To be applicable, Good
makes the following approximation
$$
{\mathcal E}_{T+1}(n_{r+1}) \approx n'_{r+1}, \qquad
\frac {T+1} T \approx 1,
$$ where $n'_r$ is the smoothed number of species seen $r$ times.  The
need for smoothing arises because at large $r$, i.e. for species that
occurred many times, the discreteness of the measurement $n_r$ within
$T$ trials becomes apparent.  Gale and Sampson provide a method, which
we used, for smoothing the data~\cite{GaleSampson:1995}.  Figures 1
and 2 in that publication make clear the need for, and effect of,
smoothing.  The result is the following estimate for the probability
of a species (configuration) that occurred $r$ times in $T$
trials~\cite[Eqs. 2 and 2']{Good:1953}
$$
p_r = \frac { r+1 } T \frac { n'_{r+1}  } { n'_r }
$$ The estimated probability of all species that occurred $r$ times,
denoted $\tilde p_r$, is
$$
\tilde p_r = \frac { r+1 } T n'_{r+1} 
$$ Summing these estimated probabilities over all
observed species~\cite[Eqs. 7,8]{Good:1953} gives
$$
\sum_{r=1}^\infty \tilde p_r = \frac 1 T \sum_{r=1}^\infty n'_{r+1} (r+1) = 
\frac 1 T \left(\sum_{r=1}^\infty n'_{r} r  - n'_1\right).
$$ If the smoothing process is performed so that Eq.~\eqref{eq:sp}
holds with $n_r$ replaced with $n'_r$, then the combined probability
of all of our observed species is given by
$$
\frac 1 T ( T - n'_1) = 1 - \frac {n'_1} T
$$
and so the estimate for the probability of the unseen species is
$$
p_0 = \frac {n'_1} T.
$$

\begin{figure}
\begin{center}
\includegraphics[height=9cm]{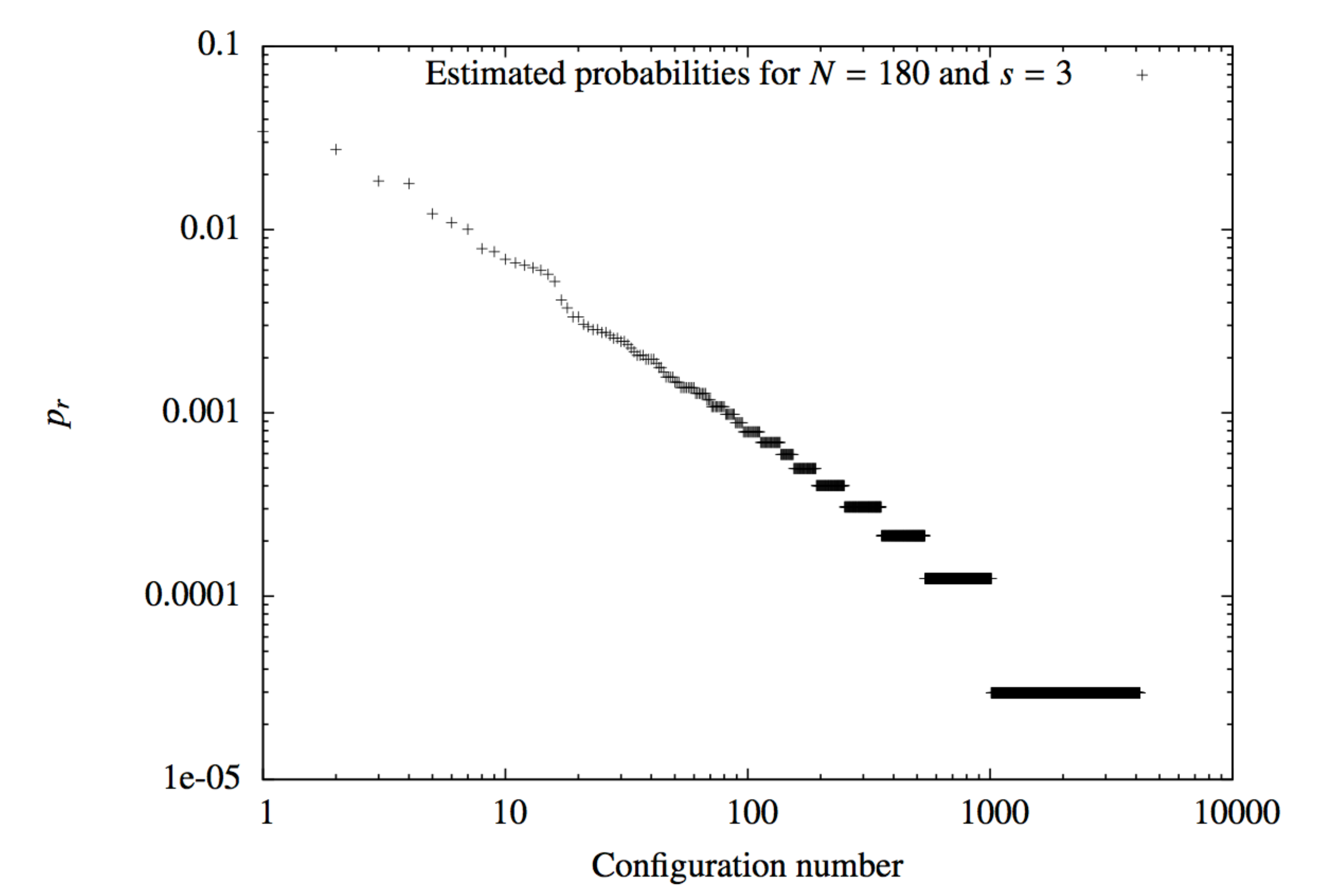}
\end{center}
\caption{\label{fig:unsmoothed} This plot shows the estimated
  probability for all of the observed configurations, ordered in
  decreasing probability.}
\end{figure}

\begin{figure}
\begin{center}
\includegraphics[height=9cm]{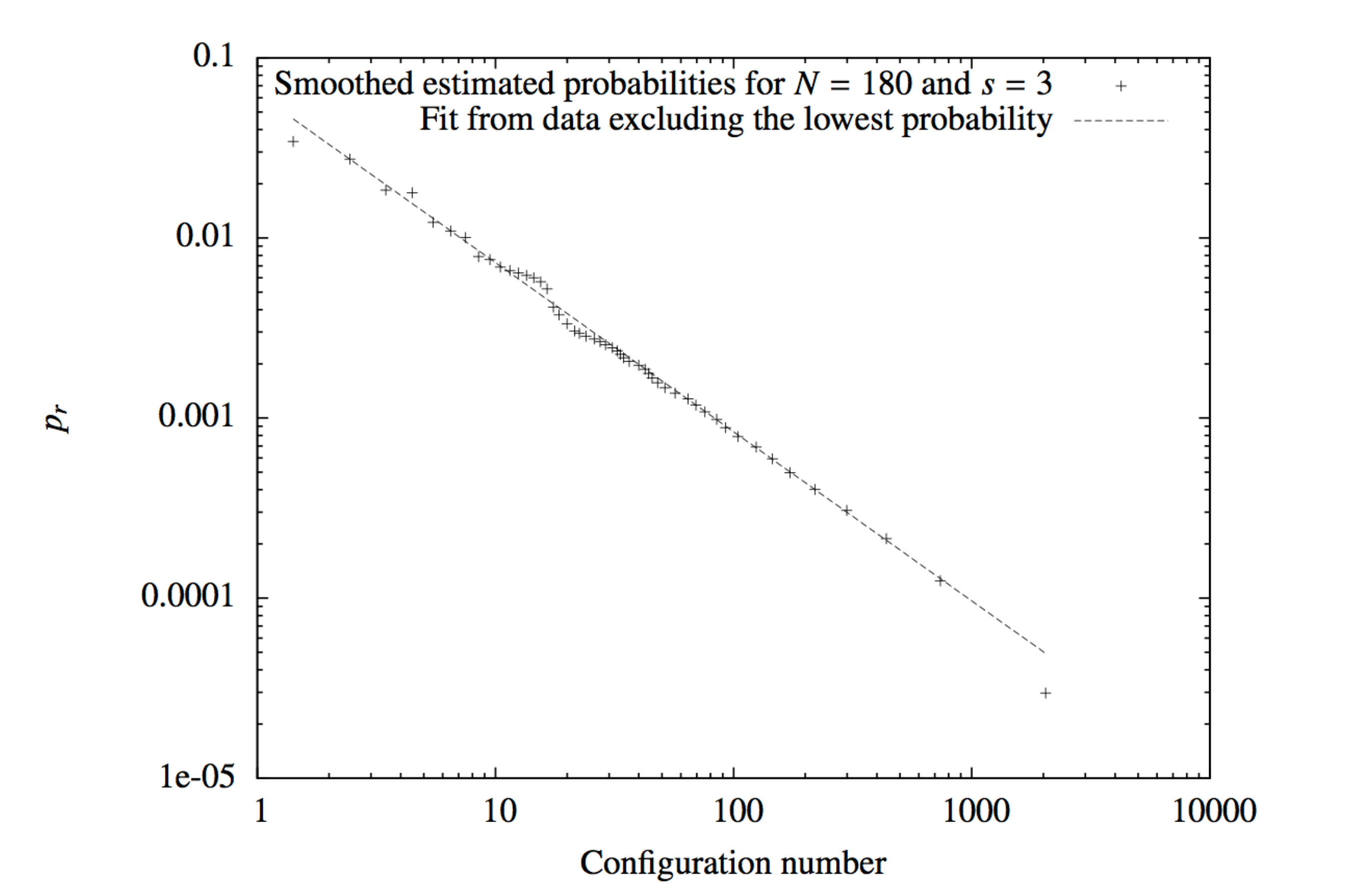}
\end{center}
\caption{\label{fig:smoothed} This plot shows the smoothed estimated
  probability for all of the observed configurations.}
\end{figure}

This method gives us estimated probabilities for each of the
configurations we have observed.  For $N=180$ and $s=3$ we show these
estimated probabilities in Figure~\ref{fig:unsmoothed}.  The
stair-step nature for the low estimated probabilities is an artifact
of the finite number of samples.  The estimated probability for a
configuration depends only on the number of times the configuration
occurred.  This number can only be $1,2,\ldots$.  Gale and Sampson's
smoothing technique addresses a similar problem, in that the observed
values of $n_r$ for large $r$ must also be integral.  We use a simple
smoothing technique where, for a given probability, we take the
geometric mean of the first and last configuration number as an
estimate for the configuration number where that probability would
occur.  This is shown in Figure~\ref{fig:smoothed}.

We fit $Ax+ \log b$ to the log-log tail of these data excluding the
point corresponding to the configurations that occurred once.  We are
operating on two assumptions here: first, the tail of the probability
distribution can be approximated $p(n) \approx b n^A$, and second,
that the last point in Figure~\ref{fig:smoothed} is not ``below'' the
fit line, as much as it is ``to the left'' of the fit line. That is to
say, for this sample, we believe that many more unseen configurations
whose probability is close to $p_1$ than there are unseen
configurations whose probability is close to or higher than $p_2$.
\begin{figure}
\begin{center}
\includegraphics[height=9cm]{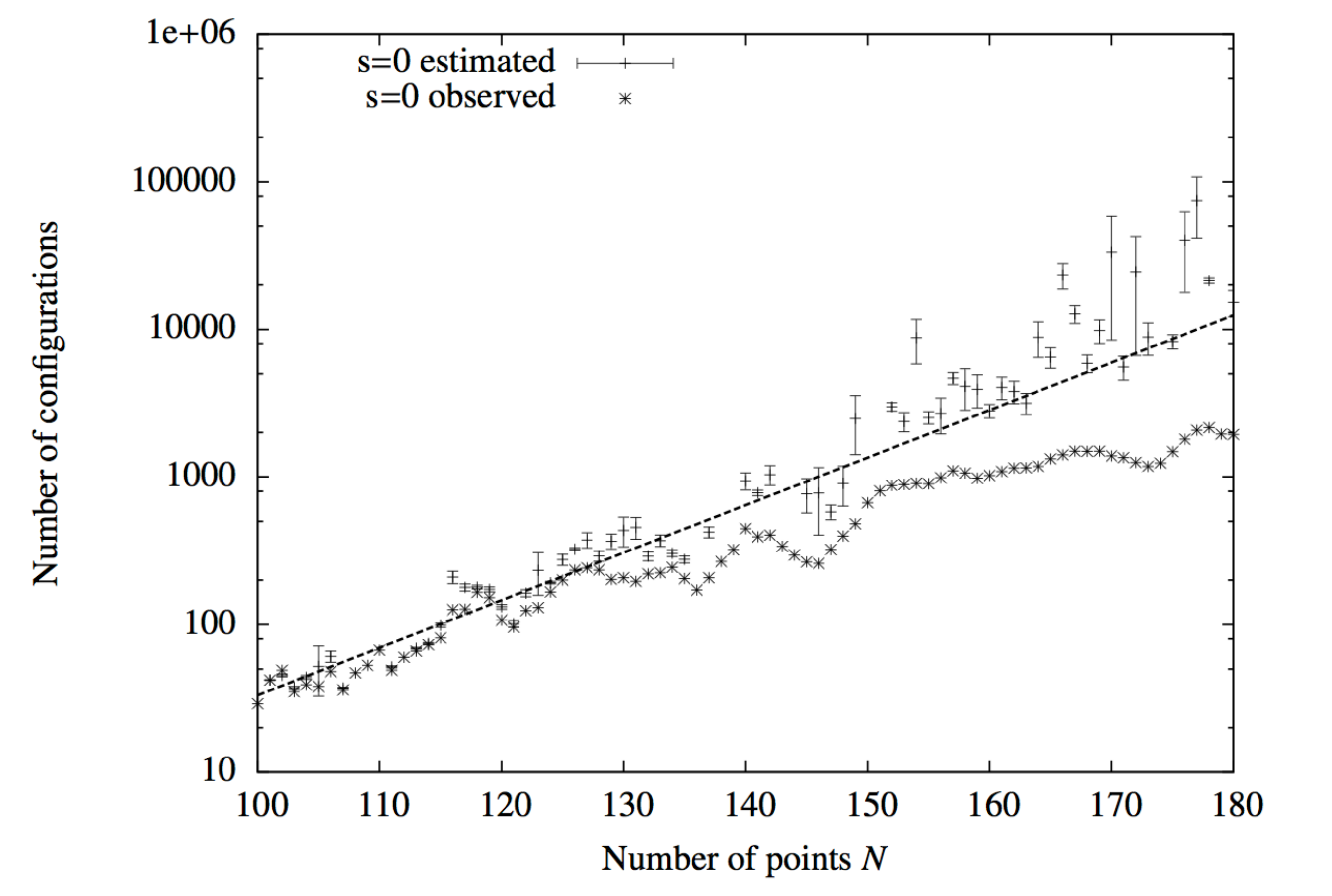}
\end{center}
\caption{\label{fig:estimated_stable_count_s0} The number of observed
  distinct stable configurations and estimates for the total number of
  distinct stable configurations as a function of $N$ for $s=0$.}
\end{figure}
\begin{figure}
\begin{center}
\includegraphics[height=9cm]{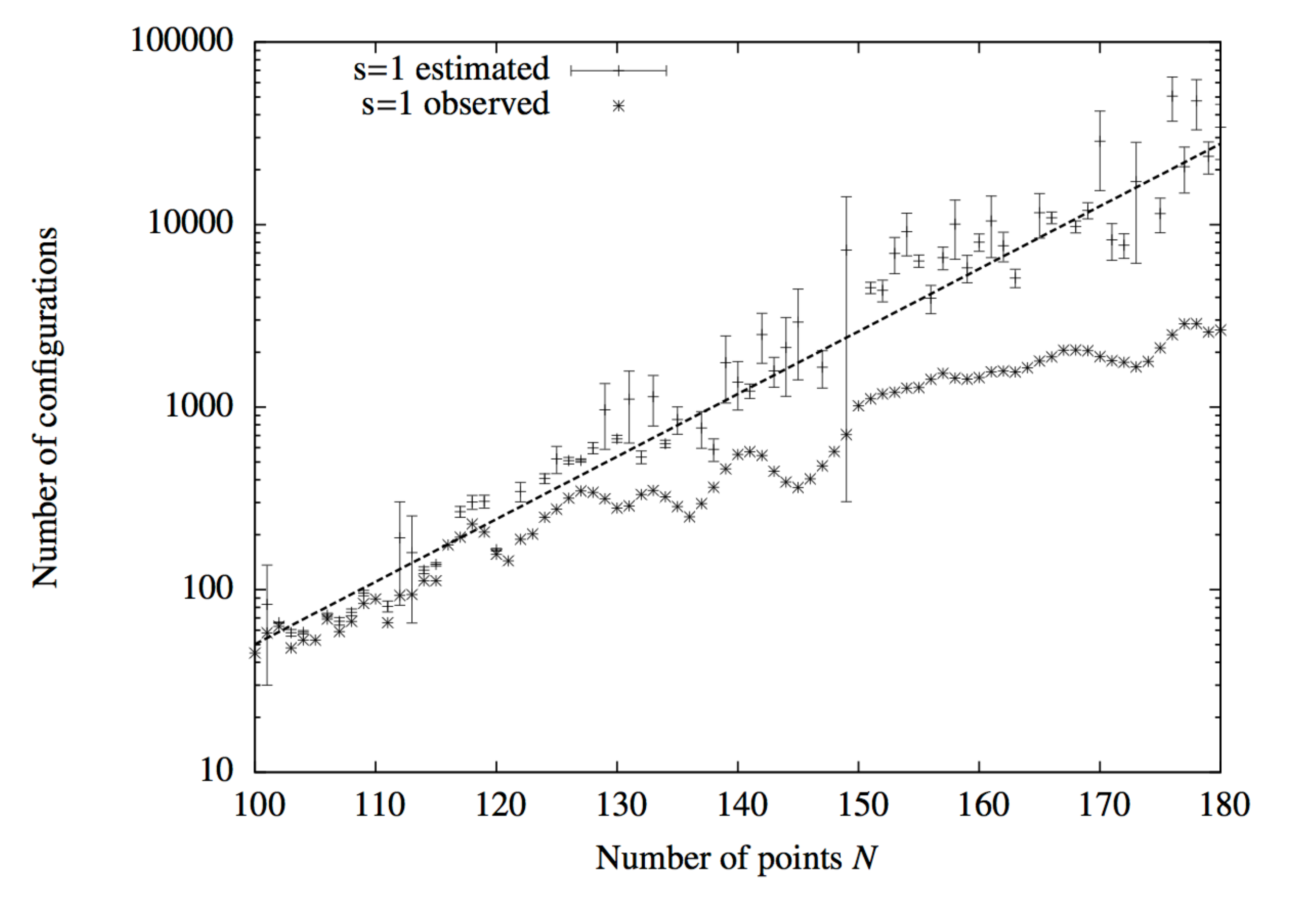}
\end{center}
\caption{\label{fig:estimated_stable_count_s1} The number of observed
  distinct stable configurations and estimates for the total number of
  distinct stable configurations as a function of $N$ for $s=1$.}
\end{figure}
\begin{figure}
\begin{center}
\includegraphics[height=9cm]{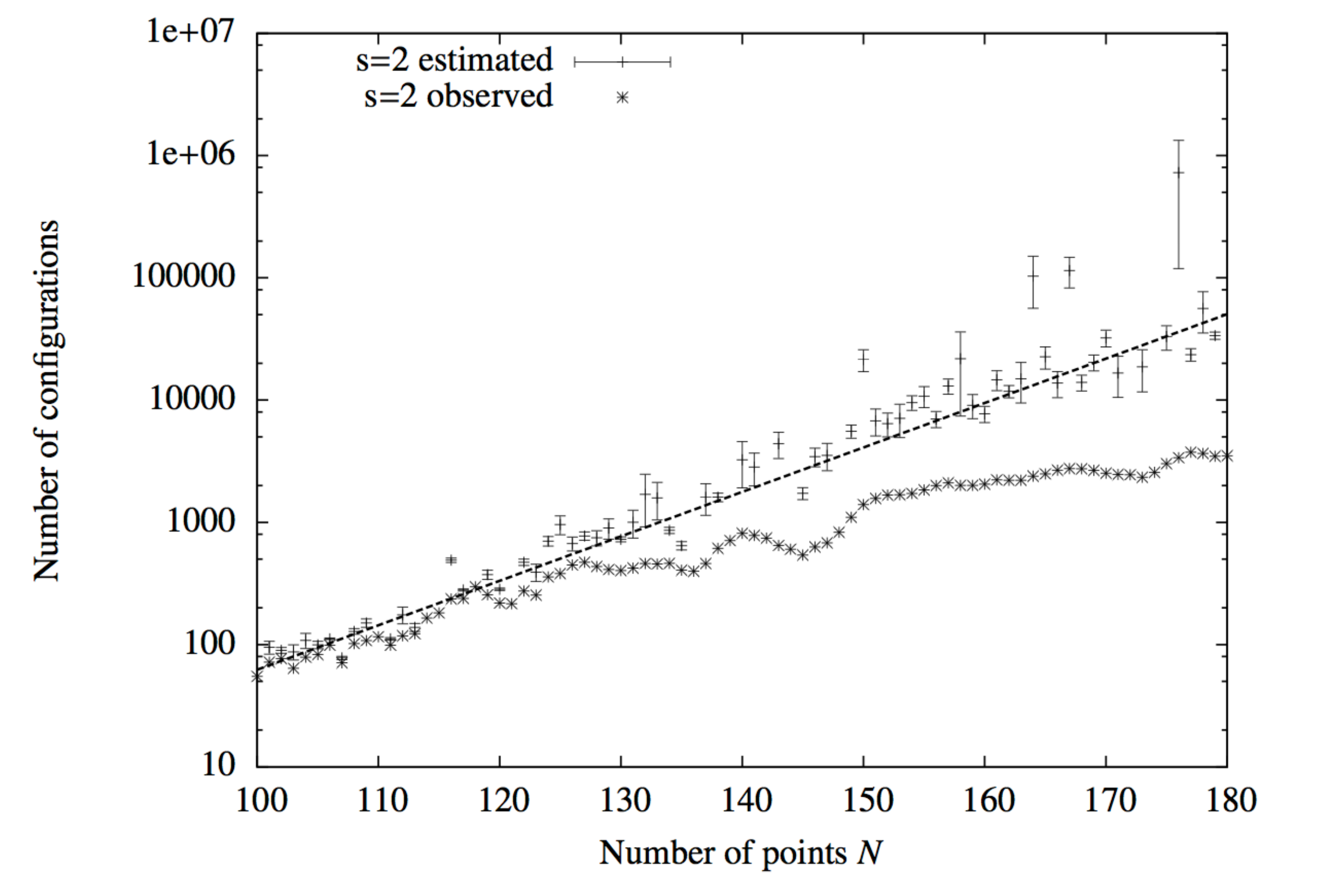}
\end{center}
\caption{\label{fig:estimated_stable_count_s2} The number of observed
  distinct stable configurations and estimates for the total number of
  distinct stable configurations as a function of $N$ for $s=2$.}
\end{figure}
\begin{figure}
\begin{center}
\includegraphics[height=9cm]{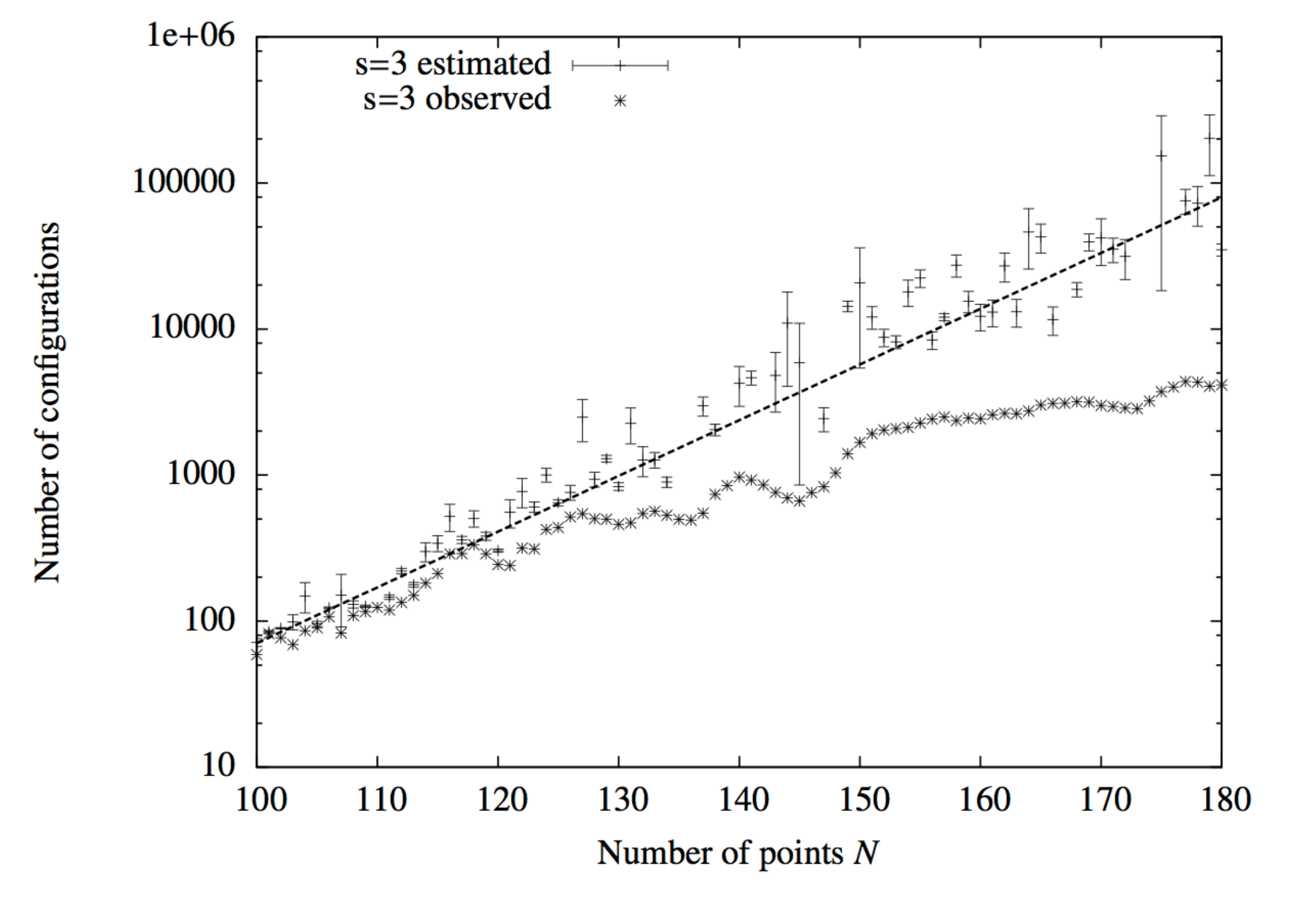}
\end{center}
\caption{\label{fig:estimated_stable_count_s3} The number of observed
  distinct stable configurations and estimates for the total number of
  distinct stable configurations as a function of $N$ for $s=3$.}
\end{figure}
Another statement of this assumption is that nearly all the unseen
configurations have probability less than $p_2$, but not necessarily
less than $p_1$.  

With this in mind we solve the following for $T_f$
\begin{equation}\label{eq:solve}
p_0 + n_1 p_1 = \int_{T_i}^{T_f} bn^A dn,
\end{equation} where $p_0$ is the estimated combined probability of all 
the species we didn't see, $p_1$ is the estimated probability for the
species we saw once, $n_1$ is the number of species that we saw once,
and $T_i$ is the number of configurations we saw at least twice.  Note
that we are only guaranteed to get a value of $T_f$ if $A\ge -1$.
When we apply this method we get an estimate for the total number of stable
configurations.  These estimates as a function of $N$ are plotted in
Figures~\ref{fig:estimated_stable_count_s0},
\ref{fig:estimated_stable_count_s1},
\ref{fig:estimated_stable_count_s2} and
\ref{fig:estimated_stable_count_s3} for $s=0,1,2$ and $3$
respectively.  In these figures we've only plotted results where the
error was less then the value itself.

If these estimates for the number of stable configurations are reasonable, and
if the growth in the number of stable configurations is exponential, then fits
from $N=100,\ldots,180$ for the number of stable configurations as a function
of $N$ and $s$ indicate that the number of stable configurations as a function
of $N$ and $s$ is given by
$$
M(N,s) = C_s \exp(e_s N),
$$
where $e_s$ and $C_s$ are given by
$$
\begin{array}{cc}
e_0 = 0.0741345 \pm 0.002804 & C_0 = \exp(-3.91164 \pm 0.3044), \\
e_1 = 0.0789298 \pm 0.00176  & C_1 = \exp(-3.97635 \pm 0.1992), \\
e_2 = 0.0836987 \pm 0.002186 & C_2 = \exp(-4.23711 \pm 0.2583), \\
e_3 = 0.0878486 \pm 0.001698 & C_3 = \exp(-4.52585 \pm 0.1882).
\end{array}
$$ 

If we perform a similar fit to the number of observed configurations
for $s=1$, as opposed to the number of estimated configurations, we
obtain $M(N,1) = (.31701 \pm .1)\,\exp(.0518 \pm .0012\,N)$, which is
similar to Erber's and Hockney's estimate of $M(N,1) = 0.382\,
\exp(0.0497N)$ noted above.  This growth is considerably slower than
the growth in the estimated number of stable configurations, which we
feel is likely closer to the actual growth in the number of stable
configurations.

\section{Conclusions}

When searching for isomorphisms between a set of configurations of
points on a sphere, the use of a simple invariant under isometry, the
discrete energy $E_s$, quickly filtered out many configurations as
not-isometric.  After this a more comprehensive invariant under
isometry, the graph of non-degenerate extremal triangulations, was
extremely effective. 

It is reasonable to express concern over the number of assumptions and
over the sensitivity of $N_f$ in Eq.~\eqref{eq:solve} to the other
parameters.  The defensible conclusion is that there are substantially
more stable configurations than those we observed, and just as
performing enough trials to observe the configuration with the lowest
energy is a formidable technical challenge, so too is finding all the
stable configurations.

\begin{acknowledgments}
The authors are grateful to Mark Ellingham for his clear explanation
of Algorithm~\ref{alg:tag}.  The authors are also grateful to the
referees for their suggested changes to the manuscript.

The work of Matthew Calef was performed under the auspices of the
National Nuclear Security Administration of the US Department of
Energy at Los Alamos National Laboratory under Contract
No. DE-AC52-06NA25396. LA-UR-14-27638

The work of Alexia Schulz is sponsored by the Assistant Secretary of
Defense for Research \& Engineering under Air Force Contract
\#FA8721-05-C-0002.  Opinions, interpretations, conclusions and
recommendations are those of the authors and are not necessarily
endorsed by the United States Government.
\end{acknowledgments}

\newcommand{\etalchar}[1]{$^{#1}$}
\providecommand{\MR}{\relax\ifhmode\unskip\space\fi MR }

\providecommand{\href}[2]{#2}

\end{document}